\pdfoutput=1
\documentclass[11pt]{article}

\usepackage[final]{acl}

\usepackage{times}
\usepackage{latexsym}
\usepackage{array}
\usepackage[T1]{fontenc}
\usepackage{microtype}

\usepackage{inconsolata}

\usepackage{graphicx}
\usepackage{booktabs, multirow}
\usepackage{enumitem}
\usepackage{makecell}
\usepackage[most]{tcolorbox}
\usepackage{fancyvrb}
\usepackage{fvextra}
\usepackage{makecell}

\newcommand{\ie}{\textit{i.e.}}
\newcommand{\eg}{\textit{e.g.}}

\newcommand{\ourbench}{UI-I2E-Bench}

\newcommand{\ourmodel}{UI-I2E-VLM}
\newcommand{\ourmethod}{UI-E2I-Synth}

\usepackage{pifont}
\newcommand{\cmark}{\ding{51}}%
\newcommand{\xmark}{\ding{55}}%
\newcommand\blfootnote[1]{\begingroup\renewcommand\thefootnote{}\footnote{#1}\addtocounter{footnote}{-1}\endgroup}

\title{\ourmethod: Advancing GUI Grounding with Large-Scale Instruction Synthesis}

\author{Xinyi Liu\textsuperscript{1,2*$\dagger$} \quad Xiaoyi Zhang\textsuperscript{1$\dagger$}  \quad Ziyun Zhang\textsuperscript{1,2*} \\
\textbf{Yan Lu$^{1}$} \\
$^1$Microsoft Research Asia \\
$^2$School of Software and Microelectronics, Peking University \\ 
}

\begin{document}
\maketitle

\blfootnote{$^{\dagger}$ Equal contribution. $^*$ Work done during the internship at Microsoft Research Asia.}

\begin{abstract}
% TODO: improve abstract for`large-scale`
Recent advancements in Large Vision-Language Models are accelerating the development of Graphical User Interface (GUI) agents that utilize human-like vision perception capabilities to enhance productivity on digital devices. Compared to approaches predicated on GUI metadata, which are platform-dependent and vulnerable to implementation variations, vision-based approaches offer broader applicability.
In this vision-based paradigm, the GUI instruction grounding, which maps user instruction to the location of corresponding element on the given screenshot, remains a critical challenge, particularly due to limited public training dataset and resource-intensive manual instruction data annotation.
In this paper, we delve into unexplored challenges in this task including element-to-screen ratio, unbalanced element type, and implicit instruction. 
To address these challenges, we introduce a large-scale data synthesis pipeline \textit{\ourmethod} for generating varying complex instruction datasets using GPT-4o instead of human annotators.
% This pipeline is designed to alleviate hallucinations and generate varying complex instructions.
Furthermore, we propose a new GUI instruction grounding benchmark \textit{\ourbench}, which is designed to address the limitations of existing benchmarks by incorporating diverse annotation aspects.
Our model, trained on the synthesized data, achieves superior performance in GUI instruction grounding, demonstrating the advancements of proposed data synthesis pipeline.
The proposed benchmark, accompanied by extensive analyses, provides practical insights for future research in GUI grounding. We will release corresponding artifacts \href{https://colmon46.github.io/i2e-bench-leaderboard/}{here}.
 % Agents with this model also exhibits strong performance on various of agent benchmark.
% The results provide solid evidence for the effectiveness and efficiency of our data synthesis pipeline.
\end{abstract}

\begin{figure}[!t]
    \centering
    \includegraphics[width=1.0\linewidth]{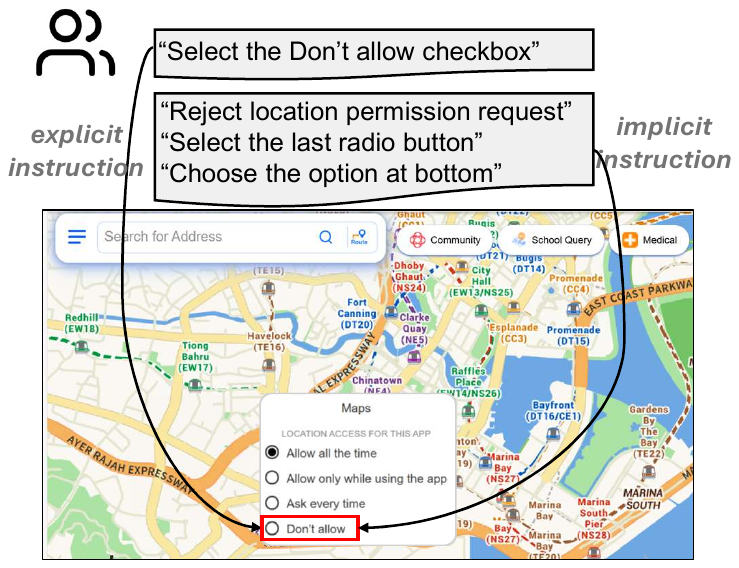}
    \caption{GUI grounding requires to localize elements in screenshots based on user instructions, with implicit instructions posing greater reasoning challenges than explicit ones.}
    \label{fig:intro}
    \vspace{1em}    

    \begin{minipage}{\linewidth}
        \centering
        \includegraphics[width=\textwidth]{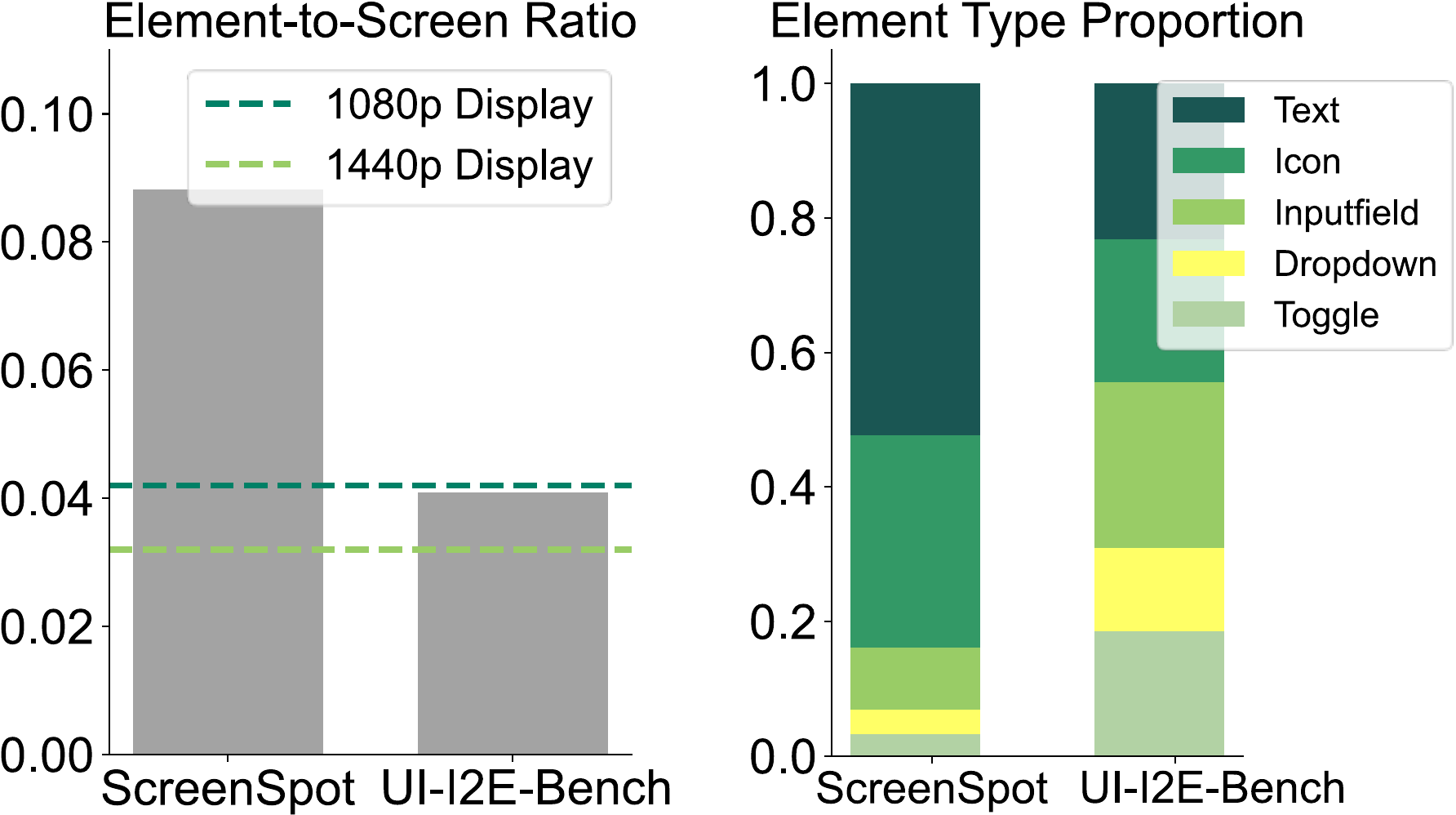}
        \label{fig:comparebench}
    \end{minipage}   
    
\vspace{-1.5em}
\caption{ \textbf{Left:} Existing benchmark's element-to-screen ratio significantly deviates from typical real-world desktop displays like 1080p and 1440p, on landscape samples. \textbf{Right:} The comparison shows Text and Icon dominates existing benchmark, leaving the rest of element types omitted.}
\vspace{-2em.}
\end{figure}

\section{Introduction}

Graphical User Interface (GUI) agents are designed to understand and automate human instruction in the digital devices, promising a future where humans are liberated from repetitive tasks, thereby enhancing operational efficiency and productivity. 
The rise of Large Language Models (LLMs)~\cite{gpt4report} made it feasible to generally understand human instructions and structured text data.
Hence, early efforts~\cite{earlyagentgur2023,webarena, earlyagentshaw2023pixels,mind2web} are made to operate GUI by understanding the text-rich metadata behind GUI, \eg, HTML.
However, some research~\cite{fu2024pixelwords,zheng2024seeact, webaim2024} has shown that GUI metadata is vulnerable and highly dependent on application developers for specific platforms. 
To overcome these limitations, more recent research~\cite{zhang2023rta,cheng2024seeclick,uground,wu2024atlas,lin2024showui} has concentrated on training Vision-Language Models (VLMs) to mimic human behavior in GUI operations, manipulating GUI elements from pixels based on user instructions.
This approach makes GUI grounding capability as the core, which maps the user instruction into the corresponding element by outputting specific coordinates.

Despite these works show satisfying progress on existing benchmark~\cite{seeclick}. Our investigation shows that there still is big gap between the existing benchmark and complicated and various GUI grounding environments. 
In this work, we thoroughly study the whole progress to build an element grounding dataset from screenshot collection to final instruction generation, revealing three main aspects that have not been fully explored previously: \textbf{(a) Element-to-screen ratio}: GUI instruction grounding necessitates higher resolution and smaller elements compared to natural scene object grounding. While previous studies~\cite{seeclick,uground} have discussed the screenshot resolution in training dataset, we argue that the key factor is the element-to-screen ratio, \ie, the size of an element relative to the screenshot. This ratio is affected by both resolution and UI zoom level. As shown in the left of Figure~\ref{fig:comparebench}, the landscape samples in existing benchmark has a larger element-to-screen ratio than typical real-world desktop displays like 1080p and 1440p, potentially leading to an overestimation of model performance.
\textbf{(b) Unbalanced element type}: Different GUI element types exhibit diverse appearances and interaction designs, also with varying frequencies of occurrence. For instance, text buttons represent the most prevalent category, whereas checkboxes are comparatively less common. Unlike text buttons, whose functionality is often explicitly conveyed by their textual labels, checkboxes typically rely on surrounding elements to define their purpose. This distinction has not been adequately addressed in previous research.
\textbf{(c) Implicit instruction}: 
In the context of GUI instruction grounding, users could convey their instructions based on their own intuitive understanding of element functionality or position, causing lack of direct correspondence with visible text on the screenshot, which we refereed as ``implicit instruction''. Figure~\ref{fig:intro} showcases the complexity of implicit instruction. The implicit instruction challenges the understanding and reasoning capability of VLMs.

To address the aforementioned challenges, we introduce a large-scale instruction grounding data synthesis pipeline, termed \textbf{E}lement-to-\textbf{I}nstruction \textbf{Synth}esis, abbreviated as \textbf{\textit{\ourmethod}}. 
This pipeline utilizes GPT-4o instead of human annotators to synthesize realistic grounding user instructions of varying complexity, which differs from the element referring expressions that previous works curated.
By the principle of divide and conquer, \textit{\ourmethod} decomposes grounding instruction synthesis into three subtasks and execute them step by step.
Initially, we collect screenshot-metadata pairs from various sources at different resolutions, then using a heuristic GUI metadata parser to extract reliable element attributes.
Next, With these high-quality attributes to alleviate hallucinations, both explicit and implicit referring expressions are prompted to generate from GPT-4o.
In the final step, GPT-4o is employed again to simulate user behavior by generating specific action parameters. These action parameters are then combined with the REs to synthesize final realistic user instructions.

To more comprehensively evaluate model performance on GUI instruction grounding, we introduce a new benchmark \textit{\textbf{\ourbench}} that includes detailed annotations curated through a combination of our synthesis pipeline and human annotators. These annotations encompass various dimensions such as element type, element-to-screen ratio, and the level of instruction implicitness. 
Through rigorous evaluation on our proposed benchmark, we demonstrate that existing VLMs are still far from being satisfactory.
To address the gap, we apply the proposed \textit{\ourmethod} to collect a synthesized GUI grounding dataset comprising 9.9M instructions. 
Then following OS-Atlas~\cite{wu2024atlas} we fine-tune two different pre-trained VLMs on the curated dataset and evaluate them.
The experimental results demonstrate that our models achieve advanced performance on both the existing benchmarks and our proposed one, with substantially less instruction data. 
Notably, our models show outstanding capabilities in comprehending implicit instructions and handling long-tailed element types. These findings validate the efficacy of our data synthesis framework for GUI grounding, while our comprehensive benchmark analysis provides practical insights for future research in this domain.

\section{Related work}

\paragraph{Autonomous GUI Agent and Vison-Language Models.}

With the development of LLMs, researchers have built autonomous agents to perform complex reasoning tasks~\cite{sumerscognitive, yang2024swe, xi2025rise}. Early works~\cite{earlyagentgur2023, webarena, earlyagentshaw2023pixels, mind2web} have initiated the development of GUI agent frameworks for webpages and mobile platforms using GUI metadata.
To address the limitations of unstable GUI metadata, \textit{ResponsibleTA}~\cite{zhang2023rta} introduced a vision-based GUI agent framework and \textit{SeeClick}~\cite{cheng2024seeclick} highlighted the importance of instruction grounding in downstream GUI agent tasks. While works like RUIG~\cite{zhang2023reinforced} improved GUI grounding via reinforcement learning, current general-purpose VLMs still lack satisfactory GUI grounding capabilities. Consequently, recent studies \cite{uground, wu2024atlas, lin2024showui} have focused on enhancing the GUI grounding abilities of VLMs. 
simultaneously, in industry, closed-source solutions~\cite{operator,computeruse} have advanced rapidly but with minimal detail disclosure.
Our work from data perspective contributes by synthesizing large-scale GUI grounding instructions and is intended to benefit the open-source community.

\paragraph{Instruction Dataset Synthesis.}
Instruction synthesis~\cite{synthding2023enhancing,synthwang2022self,synthxu2023wizardlm} has emerged as a prevalent strategy in the field of large language models due to its effectiveness in reducing the need for human effort in labeling datasets of millions of samples.
Recent works make preliminary exploration about GUI grounding data synthesis. 
\textit{SeeClick} utilizes HTML tags to generate referring expressions for webpage elements. \textit{UGround} improves on this by employing a hybrid pipeline that creates referring expressions using both external LLMs and VLMs. \textit{OS-Atlas} derives instructions from sequential UI changes but faces scalability issues due to the need of screenshot traces.
Our data synthesis pipeline differs from previous works in several key aspects, most notably in that prior studies have focused solely on element referring expressions rather than genuine user instructions. These existing approaches overlook the complex role that users play in formulating instructions and also screen-to-element ratio and long-tailed element type.

\begin{figure*}
    \centering
    \includegraphics[width=\textwidth]{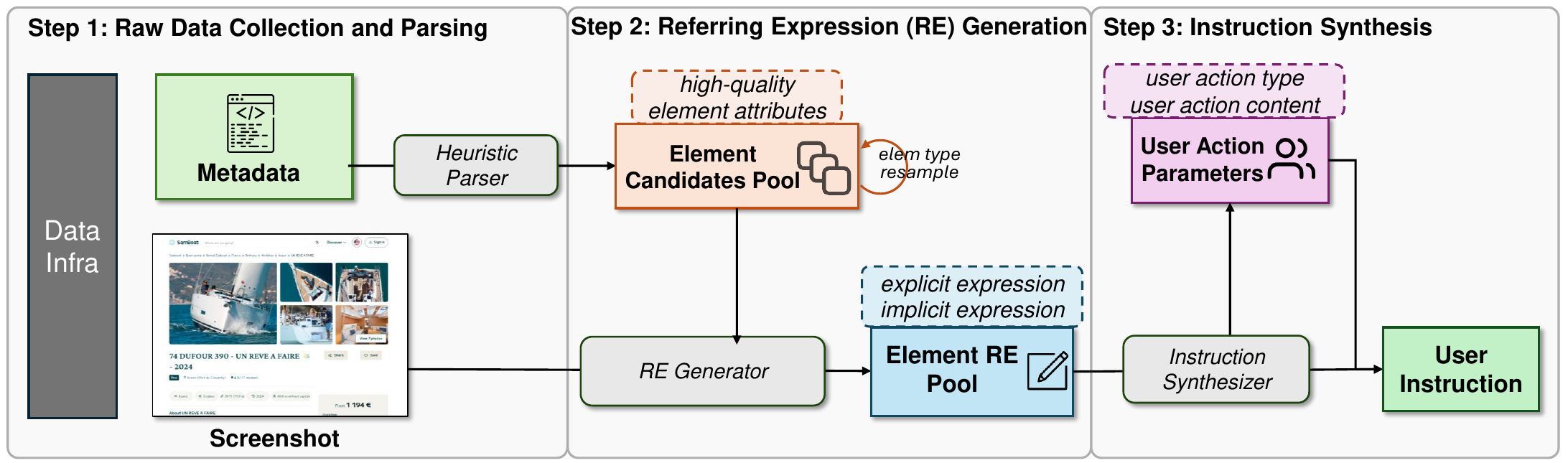}
    \caption{Overview of proposed data synthesis pipeline \ourmethod. The three steps respectively generate a high-quality pool of elements and their corresponding attributes, diverse element referring expressions, and user instructions. The output of the synthesis pipeline is format as {<screenshot, user instruction, element coordinates>}. GPT-4o is utilized in the component \textit{Referring Expression Generator} and \textit{Instruction Synthesis}.  }
    \label{fig:overview}
\end{figure*}

\section{Approach}
Model-based large-scale GUI grounding instruction synthesis is challenging for multimodal hallucinations and varying complexity of user instruction.
\textit{\ourmethod} includes three key steps: raw data collection and parsing, referring expression generation and instruction synthesis, as illustrated in Figure~\ref{fig:overview}.
These steps are executed sequentially, with each step designed to address specific challenge: extracting reliable element attributes to mitigate hallucinations, generating explicit and implicit referring expressions to diversify instruction difficulty, and synthesizing instructions with user action simulation. In the following subsections we introduce each step on details.

\subsection{Raw Data Collection and Parsing}
This component is designed to collect diverse screenshots from multiple platforms and obtain reliable element attributes from GUI metadata for the following referring expression generation.

We first build data collection infrastructure across various platforms to obtain the GUI screenshot and metadata pairs. 
For Web platform, we use the dumped webpage metadata in Common Crawl~\cite{commoncrawl} and re-render as webpage screenshots. 
For Windows platform, we select common applications, traverse the application to obtain different UI screenshots. Specifically, we start from the inital UI of a given application, parse out clickable buttons from its metadata and click these buttons to transit to other UI in this application.All the process is conducted under a virtual machine where no personal information is involved and all corresponding UIA data is recorded.
For mobile platform, we leverage the existing metadata-screenshot pairs in training set of AndroidControl~\cite{li2024androidcontrol}.
Different platforms yield different metadata formats, such as DOM~\cite{dom} for Web, UIA~\cite{msuia} for Windows, VH~\cite{androidvh} for Android.
To unify element representation from different platform metadata, we build corresponding heuristic element parser to extract only three key attributes including element type, element content and element bounding box.
For element type, we analyze the UI element representation in different metadata format and categorize them into five main element types: 
\begin{itemize}[leftmargin=*, itemsep=0pt, topsep=2pt, parsep=0pt, partopsep=0pt]
    \item \textbf{Text}: A button element with explicit text.
    \item \textbf{Inputfield}: An element requiring user input or content editing.
    \item \textbf{Dropdown}: A button that allows user input by selecting from provided options.
    \item \textbf{Icon}: A graphical button representing functionality through an image.
    \item \textbf{Toggle}: A two-state element, such as checkboxes, radio buttons, and switches.
\end{itemize}

Accurately parsing every element for each platform is extremely difficult, which means tons of rules need be constructed manually. 
However, GUI grounding data does not require perfect parsing for every element in a single screenshot. Instead, our goal is to extract sufficiently accurate element attributes from large-scale screenshot-metadata pairs.
With this goal, we build our heuristic metadata parser which contains heuristic rules to only keep high precision but leaving recall aside.

After obtaining the three key element attributes, we measure the element type distribution, adjust the distribution through resampling, and compose a balanced candidate pool for the next step.

\begin{table*}[h!]
\centering
\resizebox{\textwidth}{!}{
\begin{tabular}{ccccccc}
\hline
\textbf{Benchmark} &\textbf{\begin{tabular}[c]{@{}c@{}}Sample\\ Num\end{tabular}} &\textbf{\begin{tabular}[c]{@{}c@{}}Landscape element-\\ to-screen ratio\end{tabular}} & \textbf{\begin{tabular}[c]{@{}c@{}}Fine-grained \\ type annotation\end{tabular}} & \textbf{\begin{tabular}[c]{@{}c@{}}Min type\\ proportion\end{tabular}} & \textbf{\begin{tabular}[c]{@{}c@{}}Implicitness\\ annotation\end{tabular}} & \textbf{\begin{tabular}[c]{@{}c@{}}Implicit instruction\\ proportion\end{tabular}} \\ \hline
\textit{ScreenSpot}  & 1,272 & 0.088   & \xmark   & 3.21\%$^{\dag}$   & \xmark   & -    \\ 
\hline
Our~\textit{\ourbench} & 1,477 &\textbf{0.042} & \textbf{\cmark}   & \textbf{12.34\%}   & \textbf{\cmark}        & \textbf{63.03\%}  \\ 
\hline
\end{tabular}
}
\vspace{-5pt}
\caption{Comparison between proposed benchmark and existing benchmark. Screen-to-element ratio is calculated as the ratio of the square roots of the areas of the box and the image. $^{\dag}$ denotes we re-annotate ScreenSpot-Web with our fine-grained element type and calculate the number. }
\label{tab:bench1}
\vspace{-7pt}
\end{table*}

\subsection{Referring Expression Generation}
The Referring Expression Generation step is designed to enable exsiting Large VLM to generate expressions that are both reliable and diverse in difficulty. 
The element referring expressions (REs) denote a element description from a specific perspective, which is independent from user action.
Generating elements and their descriptions from scratch can be challenging for Large VLMs, as they often struggle to produce accurate coordinates and are prone to hallucinations~\cite{zheng2024seeact} under high-resolution images.
A straightforward method of generating element REs is using Set-of-Marks~\cite{gpt4vsom}  to caption every marked element.
However, this method also tends to result in severe hallucinations. To address this issue, we propose a attribute-enhanced style of RE generation.
We supply a list containing the element types and element content obtained from the previous step, with Set-of-Marks screenshots as context.
We then leverage GPT-4o as the Large VLM to first generate a full description that explains the function of each element and the expected outcome when interacting with it.
Next, GPT-4o is prompted with generating two types of REs: explicit RE and implicit RE. 
We define explicit RE as directly referring to the obvious features of the element, while implicit RE refers to the element by describing the semantic function or its relationship with nearby elements, thus avoiding the obvious visible features.
The generated explicit and implicit REs collectively form the element RE pool, which provides user action object for the final step.

\subsection{Instruction Synthesis}
The obtained referring expressions are only descriptions about element, omitting the user role in the progress. 
User instruction implies intention from user, which can be directly or implicitly related to the element.
When directly asked to generate user instruction, we observed that Large VLM trends to generate generally instruction from a role of assistant, such as ``Fill your details in the input field'' or ``Click to check your profile'', even we ask it to generate as a computer user's first perspective view.
This phenomenon may arise because the output message role in the Large VLM is set to ``assistant'', preventing it from easily simulating the user role.
Hence, here we propose to a parametrization way to generate user instruction.
We decompose the wanted instruction into three parameters, user action type, user action content and element object.
A prompt is utilized to instruct GPT-4o to simulate a user interacting with the current application, by generate specific user actions and content. 
Combining previous element referring expressions as the element object, GPT-4o is further instructed with synthesizing the final instruction with all these action parameters. 
To maintain diversity in the generated instructions, both explicit and implicit element referring expressions are employed to create distinct user instructions.
To this end, the output of the whole synthesis pipeline is format as \textit{<screenshot, user instruction, element coordinates>}.
% 

% In this component, we convert the generated referring expressions into user instruction by user action simulation.
With this pipeline, we synthesize a large-scale grounding instruction data to train a model to demonstrate our advancement. Dataset details are explained in Section~\ref{sec:trainingdata}.

\section{\textit{\ourbench}: A Comprehensive Grounding Benchmark}
\label{sec:ourbench}
As shown in Figure~\ref{fig:comparebench}, the existing benchmark ScreenSpot exhibits a lower level of difficulty in element-to-screen ratio. Additionally, it lacks detailed element annotations, such as element type and implicitness, which are crucial for assessing the element grounding capabilities of models. To address these limitations, we introduce \textit{\ourbench}, developed through a semi-automated workflow to provide a more comprehensive evaluation.
We first apply our data synthesis pipeline into data collected from multiple platforms, then we sampled the instructions by element-to-screen ratio and element type.
We manually review the validness of parsed element and correctness of generated instruction to obtain the final \textit{\ourbench}, which includes 1477 grounding instructions from Web, Windows and Android.
The further statistical data and comparisons with \textit{ScreenSpot} are included in Table~\ref{tab:bench1}. 
Our benchmark offers more comprehensive annotations, a lower element-to-screen ratio, and a higher proportion of implicit instructions. These features facilitate in-depth performance diagnostics in challenging GUI instruction grounding cases.

% \begin{table*}[h]
% \centering
% \resizebox{0.8\textwidth}{!}
% {
% \centering
% \begin{tabular}{lllrrrrrrrr}\toprule
% \multirow{2}{*}{\textbf{Model}} &\multirow{2}{*}{\textbf{Size}} &\multirow{2}{*}{\textbf{\#Train}} &\multicolumn{2}{c}{\textbf{Mobile}} &\multicolumn{2}{c}{\textbf{Desktop}} &\multicolumn{2}{c}{\textbf{Web}} &\multirow{2}{*}{\textbf{Avg.}} \\
% \cmidrule(lr){4-5} \cmidrule(lr){6-7} \cmidrule(lr){8-9}
% & & &Text &Icon &Text &Icon &Text &Icon & \\\midrule
% SeeClick & 9.6B & 364K & {78.0} &52.0 &72.2 &30.0 &55.7 &32.5 &53.4 \\
% % SeeClick & 9.6B & 364K & {78.0} &52.0 &72.2 &30.0 &55.7 &32.5 & 55.7 \\
% OmniParser+GPT-4o & - & - & 87.4 & 67.9 & 93.2 & 60.9 & 77.8 & 48.3 & 73.9\\
% UGround & 7B & 1.3M & 82.8 & 60.3 & 82.5 & 63.6 & 80.4 & 70.4 & 73.3\\
% % UGround & 7B & 1.3M & 82.8 & 60.3 & 82.5 & 63.6 & 80.4 & 70.4 & 74.1\\
% % OS-ATLAS-4B~\cite{} & 4B & 2.2M & 85.7 & 58.5 & 72.2 & 45.7 & 82.6 & 63.1 & 70.1\\
% OS-ATLAS-4B~\cite{} & 4B & 2.2M & 85.7 & 58.5 & 72.2 & 45.7 & 82.6 & 63.1 & 68.0\\
% % OS-ATLAS-7B~\cite{} & 7B & 2.2M & 93.0 & 72.9 & 91.8 & 62.9 & 90.9 & 74.3 & 82.5 \\
% OS-ATLAS-7B~\cite{} & 7B & 2.2M & 93.0 & 72.9 & 91.8 & 62.9 & 90.9 & 74.3 & 81.0 \\
% \midrule
% \ourmodel-4B & {4B} & 671K & 88.3 & 62.5 & 86.6 & 58.6 & 78.3 & 62.1 & 72.7 \\
% \ourmodel-7B & {7B} & 1.7M & 91.6 & 77.3 & 91.2 & 75.7 & 84.8 & 68.5 & \textbf{81.5}\\
% \bottomrule
% \end{tabular}
% }
% \caption{
% GUI grounding accuracy comparison on Screenspot. \#Train denotes the screenshot number in training data. 
% }
% \label{tab:screenspot}
% \end{table*}

\begin{table*}[h]
\centering
\resizebox{\textwidth}{!}
{
\centering
\begin{tabular}{lllccccccccc}
\toprule
\multirow{2}{*}{\textbf{Model}} & \multirow{2}{*}{\textbf{Size}} & \multirow{2}{*}{\textbf{\#Train}} & \multicolumn{4}{c}{\textbf{ScreenSpot}} & \multicolumn{3}{c}{\textbf{\ourbench}} & \multirow{2}{*}{\textbf{ScreenSpot-Pro}} & \multirow{2}{*}{\textbf{\textit{Avg.$^{*}$}}}\\
\cmidrule(lr){4-7} \cmidrule(lr){8-10} 
 & & & \textbf{Mobile} & \textbf{Web} & \textbf{Desktop} & \textbf{Avg.} & \textbf{Explicit} & \textbf{Implicit} & \textbf{Avg.} & & \\
\midrule
InternVL2-4B & {4B} & - & 7.2 & 0.5 & 4.5 & 4.2 & 1.4 & 0.5 & 0.9 & 0.3 & \textit{1.8}\\
Qwen2-VL-7B & {7B} & - & 51.3 & 27.7 & 49.1 & 42.6 & 53.8 & 45.6 & 48.7 & 1.6 & \textit{31.0}\\
OmniParser & - & - & 78.5 & 63.9 & 79.7 & 73.9 & 54.3 & 52.4 & 53.1 & 8.3 & \textit{45.1} \\
Seeclick & {9.6B} & 1.0M & 66.1 & 44.7 & 54.5 & 55.8 & 37.1 & 19.9 & 26.4 & 1.1 & \textit{27.8}\\
UGround & {7B} & 9.7M & 72.5 & 75.7 & 74.6 & 74.1 & 65.8 & 47.1 & 54.2 & 16.5 & \textit{48.3}\\
ShowUI & {2B} & 2.7M & 84.6 & 73.2 & 69.9 & 76.8 & 51.3 & 35.6 & 41.5 & 7.7 & \textit{42.0}\\
OS-Atlas-4B & {4B} & 13.6M & 73.3 & 73.4 & 61.1 & 70.1 & 51.5 & 39.9 & 44.3 & 3.7 & \textit{39.4}\\
OS-Atlas-7B & {7B} & 13.6M & 83.8 & \textbf{83.1} & 79.7 & 82.5 & 63.2 & 55.8 & 58.6 & 18.9 & \textit{53.3}\\
\midrule
\ourmodel-4B & {4B} & 9.9M & 70.3 & 70.9 & 70.1 & 70.4 & 61.9 & 48.3 & 53.4 & 12.2 & \textit{45.3}\\
\ourmodel-7B & {7B} & 9.9M & \textbf{86.5} & 78.0 & \textbf{82.6} & \textbf{82.5} & \textbf{72.0} & \textbf{67.9} & \textbf{69.5} & \textbf{23.6} & \textit{\textbf{58.5}}\\
\bottomrule
\end{tabular}
}
\caption{
Results on GUI grounding benchmarks. \#Train denotes the instructions number in training data. We calculate the average accuracy as correct samples divided by total samples. \textit{Avg.$^{*}$} denotes the arithmetic mean of average accuracy across three benchmarks.
}
\label{tab:screenspot}
\end{table*} 
% \begin{table*}[h!]
% \centering
% \resizebox{0.9\textwidth}{!}
% {
% \centering
% \begin{tabular}{llllcccccc}
% \toprule
% \multirow{2}{*}{\textbf{Model}} & \multirow{2}{*}{\textbf{Size}} &\multirow{2}{*}{\textbf{\#Train}} & \multicolumn{3}{c}{\textbf{Platform}} & \multicolumn{2}{c}{\textbf{Implicitness}} & \multirow{2}{*}{\textbf{Avg.}} \\
% \cmidrule(lr){4-6} \cmidrule(lr){7-8}
%  & & & \textbf{Web} & \textbf{Desktop} & \textbf{Mobile} & \textbf{Implicit} & \textbf{Explicit} & \\
% \midrule
% Seeclick & 9.6B & 364K & &  &  & &  & \\
% OmniParser+GPT-4o &-&- & 30.83 & 45.47 & 67.57 &  &  & \\
% UGround  & 7B & 1.3M & 52.96 & 44.32 & 61.84 & 47.06 & 65.83 & 54.16 \\
% % OS-ATLAS-4B    & 54.62 & 20.67  & 61.94 & 37.90   & 57.08   & 44.99 \\
% % OS-ATLAS-7B    & 63.03 & 44.22  & \textbf{70.59} & 53.53   & 69.58   & 59.46 \\
% \midrule
% \ourmodel-4B   & {4B} & 671K & 73.52 & 56.65 & 65.67 & 58.50 & 63.89  & 63.85 \\
% \ourmodel-7B  & {7B} & 1.7M & 62.45 & 62.43 & 76.31 & 66.96 & 73.57 & 69.06\\
% \bottomrule
% \end{tabular}
% }
% \caption{GUI grounding accuracy comparison on \ourbench{}. \#Train denotes the number of screenshots in training data. Implicit and Explicit indicate implicit instruction and explicit instruction. We calculate the Avg. as correct samples divided by total samples.}
% \vspace{-5pt}
% \label{tab:ourbench}
% \end{table*}

\begin{table*}[h!]
\centering
\resizebox{0.89\textwidth}{!}
{
\centering
\begin{tabular}{lllcccccccc}
\toprule
\multirow{2}{*}{\textbf{Model}} & \multirow{2}{*}{\textbf{Size}} &\multirow{2}{*}{\textbf{\#Train}} & \multicolumn{3}{c}{\textbf{Platform}} & \multicolumn{5}{c}{\textbf{Element Type}} \\
\cmidrule(lr){4-6} \cmidrule(lr){7-11}
 & & & \textbf{Web} & \textbf{Desktop} & \textbf{Mobile} & \textbf{Button} & \textbf{Icon} & \textbf{Dropdown} & \textbf{Input} & \textbf{Toggle} \\
\midrule
% InternVL2-4B & {4B} & - & 0.8 & 1.0 & 0.9 & 0.4 & 0.4 & 0.0 & 3.4 & 0.0 \\
% Qwen2VL-7B & {7B} & - & 44.7 & 38.9 & 57.3 & 56.1 & 44.8 & 74.7 & 60.7 & 24.8 \\
OmniParser & - & - & 30.8 & 45.5 & 67.6 & 68.4 & 60.5 & 65.9 & 58.9 & 26.9 \\
Seeclick & {9.6B} & 1.0M & 18.2 & 15.8 & 37.2 & 31.6 & 26.2 & 22.5 & 29.6 & 22.1 \\
Uground & {7B} & 9.7M & 53.0 & 44.3 & 61.8 & 57.3 & 49.7 & 76.4 & 64.2 & 37.0 \\
ShowUI & {2B} & 2.7M & 29.6 & 30.4 & 53.9 & 52.0 & 44.1 & 51.1 & 52.8 & 18.9 \\
OS-Atlas-4B & {4B} & 13.6M & 54.6 & 19.9 & 58.6 & 43.5 & 44.1 & 46.6 & 46.3 & 42.2 \\
OS-Atlas-7B & {7B} & 13.6M & 52.2 & 48.9 & 68.1 & 69.1 & 58.7 & 80.3 & 70.1 & 32.3 \\
\midrule
\ourmodel-4B & {4B} & 9.9M & 60.9 & 38.9 & 61.4 & 54.3 & 50.0 & 61.2 & 68.6 & 39.0 \\
\ourmodel-7B & {7B} & 9.9M & \textbf{62.1} & \textbf{64.0} & \textbf{76.2} & \textbf{77.0} & \textbf{68.2} & \textbf{84.8} & \textbf{86.2} & \textbf{44.4} \\
\bottomrule
\end{tabular}
}
\caption{Detailed results on \ourbench{}. \#Train denotes the instruction number in training data.}
\vspace{-5pt}
\label{tab:ourbench}
\end{table*}
\vspace{-5pt}

\section{Experiments}
To demonstrate the effectiveness of our grounding instruction synthesis pipeline, we utilize \ourmethod{} to collect a GUI grounding training dataset and fine-tune a pre-trained VLM to develop \ourmodel{}. This section is structured as follows:
First, we provide details on the curation of the training dataset and model training. 
Next, we introduce various baselines and evaluate the performance of \ourmodel{} against these baselines on two GUI instruction grounding benchmarks. We analyze the results on \ourbench{} to reveal the shortcomings of current VLMs across multiple aspects.
We then integrate \ourmodel{} with the GPT-4o planner and assess its performance on the live GUI agent benchmark OSWorld~\cite{osworld}.

\subsection{Training Details}

\paragraph{Training data curation.}
\label{sec:trainingdata}
In this subsection, we briefly describe the final curated dataset leveraging the \textit{\ourmethod} pipeline while leaving more details in our supplementary materials.% TODO 这里需要具体引用吗
We apply our pipeline mainly on three platforms including Web, Windows and Android to derive this training dataset.
The Web is our main data source due to the variety of layouts and design styles across websites, as well as the extensive quantity of webpages available in Common Crawl. We start by extracting webpages from the top 500k domains with the highest traffic for three webpages per domain. Next, we filter out non-English and error state webpages, resulting in final 724,839 webpages. These webpages are re-rendered at seven different resolutions, including one mobile resolution to simulate mobile scenarios and left six resolutions for landscape desktops.
For Windows platform, we select 90 windows apps to totally collect 15K screenshot-metadata pairs.
For Android platform, we leverage existing screenshot-metadata pairs in training set of AndroidControl~\cite{li2024androidcontrol}.
Then we run the \textit{\ourmethod} pipeline on all the collected screenshot-metadata pairs to derive our synthsized dataset. Combined wuth the existing mobile GUI dataset MOTIF~\cite{burns2022motifvln} and WidgetCaption~\cite{li2020widget} for grounding purpose, we final training dataset includes 1.6M screenshots and 9.5M instructions. 
The detailed training data statistics are provided in Appendix\ref{appendix:statistic}.
Our data has passed our ethics review where personal information has been examined and removed.

\paragraph{Base models.} Following OS-Atlas~\cite{wu2024atlas}, We consider InternVL2-4B~\cite{chen2023internvl} and Qwen2-VL-7B~\cite{qwen2vl} as our base models. The pretraining data of Qwen2-VL-7B incorporate GUI screenshots, whereas InternVL2 does not. Both models support multi-resolution image processing. We name our models as UI-I2E-VLM-4B and UI-I2E-VLM-7B, with training details provided in Appendix~\ref{appendix:training_details}.

\subsection{GUI Instruction Grounding Evaluation}
In this section, we compare~\ourmodel{} with OmniParser~\cite{omniparser}, SeeClick~\cite{seeclick}, UGround~\cite{uground}, ShowUI~\cite{lin2024showui} and OS-Atlas~\cite{wu2024atlas} on multiple GUI grounding benchmarks. OmniParser is a screen parsing tool containing a series of models. We combine OmniParser with GPT-4o to perform GUI grounding as its original paper suggests.
SeeClick, UGround, ShowUI and OS-Atlas are VLMs fine-tuned for GUI instruction grounding tasks.
Following their inference setting, we prompt SeeClick, UGround and ShowUI to generate (x, y) coordinates. For OS-Atlas and~\ourmodel{}, we prompt the model to generate a bounding box and use its center point as the predicted coordinate.

For the evaluation benchmarks, we not only include the proposed benchmark but also the popular ScreenSpot even a concurrent benchmark ScreenSpot-Pro. ScreenSpot is a popular GUI grounding benchmark consisting of 1,272 cross-platform samples. ScreenSpot-Pro is a very recent benchmark focusing on professional desktop applications with high resolution.  We use accuracy as the evaluation metric, where a prediction is considered as a correct one if the coordinate falls within the ground-truth bounding box.

\paragraph{Main result comparison.} 
We present a comprehensive performance comparison across the aforementioned three benchmarks to evaluate the models' cross-domain generalization capabilities. As demonstrated in Table~\ref{tab:screenspot}, \ourmodel-7B achieves superior performance across all benchmarks, surpassing the previous state-of-the-art model, OS-Atlas-7B, with a 9.7\% relative improvement in average performance. Notably, the robust improvement on all the three distinct benchmarks is attained with a training set containing only 72\% of the instruction quantity used in OS-Atlas, which makes our curated data quality more impressive.
For our \ourmodel-4B, despite its base model being pre-trained without GUI-specific data, still achieves impressive performance compared to the models of similar parameter size. These findings highlight the effectiveness of our data synthesis pipeline.

By comparing the numerical results between ScreenSpot and \ourbench, our findings indicate that the reported progress in GUI grounding capabilities has been substantially overestimated. A notable example is ShowUI, which demonstrates seemingly impressive performance metrics on ScreenSpot despite utilizing significantly fewer model parameters and training data.
However, its performance drops markedly when evaluated on \ourbench{} and ScreenSpot-Pro, revealing critical limitations in the complexity and representativeness of the original ScreenSpot benchmark.

\paragraph{Teardown diagnostic on \ourbench.}

Leveraging the diverse annotation aspects provided in \ourbench{}, we conducted a detailed teardown analysis of model performance on \ourbench{}. We 
report the detailed performance on the three dimensions separately in Table~\ref{tab:screenspot} and ~\ref{tab:ourbench}.
Compared to the explicit instruction category, we observe more significant improvement happening to the implicit ones. The most outstanding model, OS-Atlas-7B falls behind of us for 12.1 percentage on this dimension, which means previous works underrate on the complexity of instructions.
Another interesting observation is that OmniParser shows a competitive result on implicit instruction with understanding abilities of GPT-4o. However, it falls behind on simpler explicit instructions.
It suggests that the primary bottleneck of OmniParser lies in localizing smaller and long-tailed element. 

\begin{figure}[t]
    \centering
    \includegraphics[width=\columnwidth]{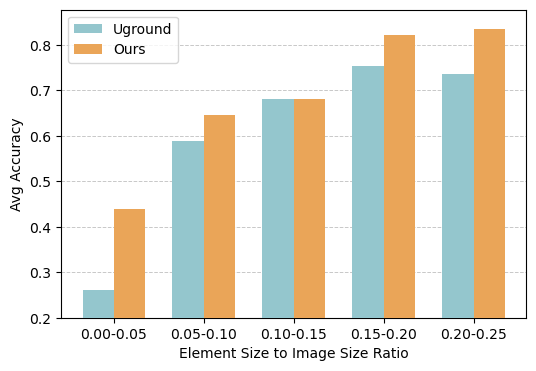}
    \caption{Accuracy comparison by the proportion of element size to image size, evaluated on \ourbench.}
    \vspace{-9pt}
    \label{fig:element_screen_ratio}
\end{figure}

To further study the impact of element-to-screen ratio in GUI grounding, we analyze the accuracy across different ranges of the element-to-screen ratio between our \ourmodel-4B and UGround, as shown in Figure~\ref{fig:element_screen_ratio}.
Grounding accuracy drops as the element size ratio decreases, highlighting the importance of benchmarks that emphasize smaller elements and higher-resolution images. Benefiting from our training data and larger number of input image tokens, \ourmodel~achieves better performance on small elements.

In element type analysis of Table~\ref{tab:ourbench}, we observe a significant performance gap in the Icon and Inputfield types compared to previous models, indicating that prior work has largely overlooked these long-tail categories. It supports our data synthesis pipeline to use a relatively balanced distribution of element types when curating the training dataset. 

\begin{table}[t]\centering
\resizebox{\columnwidth}{!}
{
\begin{tabular}{lcccccccc}\toprule
\multirow{2}{*}{\textbf{Model}} &\multicolumn{2}{c}{\textbf{Mobile}} &\multicolumn{2}{c}{\textbf{Desktop}} &\multicolumn{2}{c}{\textbf{Web}} &\multirow{2}{*}{\textbf{Avg.}} \\
\cmidrule(lr){2-3} \cmidrule(lr){4-5} \cmidrule(lr){6-7}
& Text &Icon &Text &Icon &Text &Icon & \\\midrule
OS-Atlas-Web & 35.2 & 24.9 & 67.0 & 30.7 & 70.9 & 39.8 & 44.9 \\
\ourmethod-Web & \textbf{89.0} & 34.1 & \textbf{93.3} & \textbf{42.9} & \textbf{79.6} & \textbf{44.7} & \textbf{65.8} \\
~~ - Inst. Synth. & 85.4 & 20.5 & 84.0 & 18.6 & 69.6 & 30.1 & 54.3($\downarrow 11.5$) \\
~~ - GPT-4o & 58.6 & \textbf{37.1} & 52.1 & 37.9 & 46.1 & 44.2 & 46.9($\downarrow 18.9$) \\

\bottomrule
\end{tabular}
}
\caption{
Data curation ablation experiment on ScreenSpot. Different 500k instruction data is sampled to separately fine-tune and evaluate the same model.
}
% \vspace{-8pt}
\label{tab:curation_ablation}
\end{table}

\begin{figure*}
    \centering
    \includegraphics[width=0.8\textwidth]{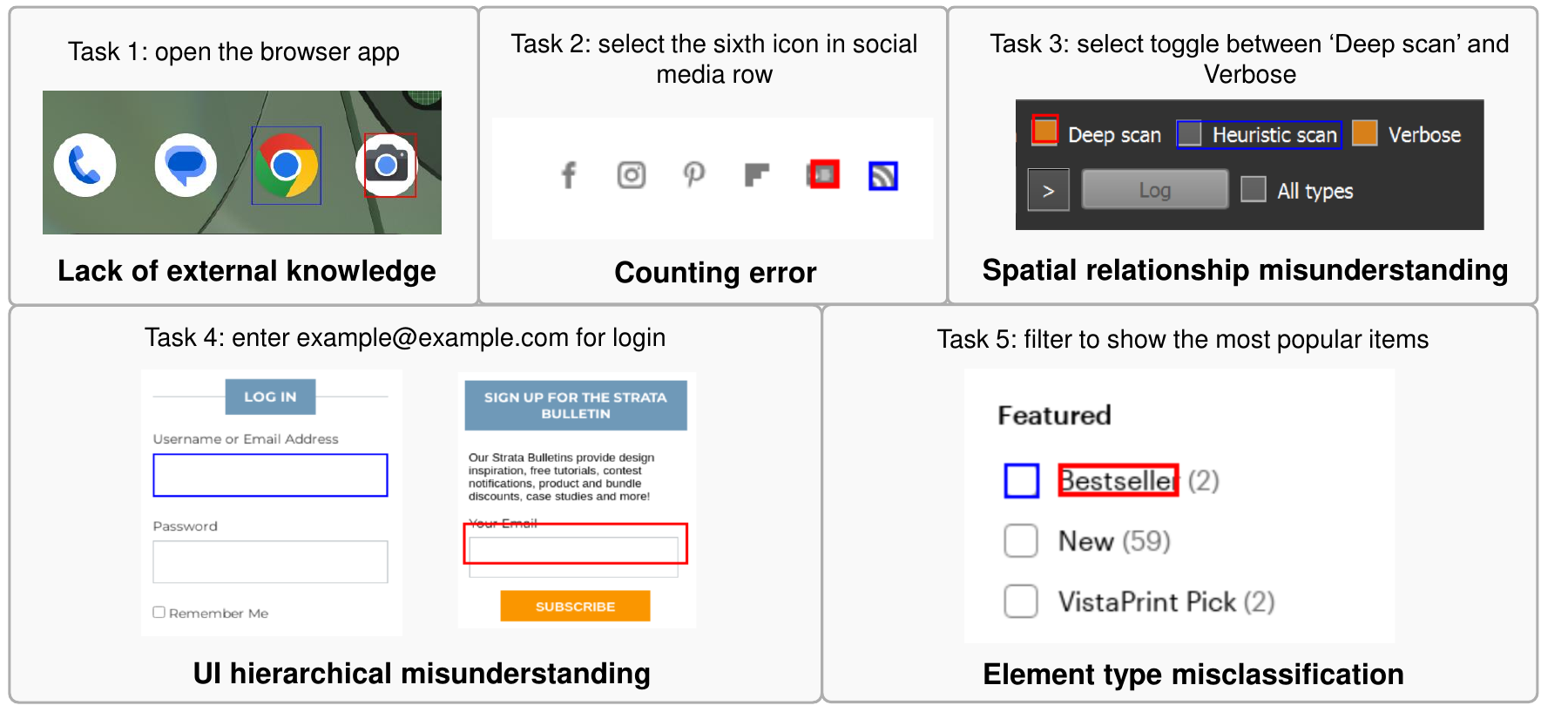}
    \caption{Typical failure cases in \ourbench{}. Blue box denotes the correct elements while red ones are wrong prediction from our model.}
    \vspace{-7pt}
    \label{failurecase}
\end{figure*}

\paragraph{Data curation ablation study.}
To fairly evaluate the quality of our curated dataset and analyze components in our data synthesis pipeline, we compare our synthesized data with public released data from OS-Atlas~\cite{wu2024atlas} and variations from our own synthesis pipeline. 
Since OS-Atlas only constructs data on the web at a large scale, we also curate data on web domain.
The comparison includes: (1) \textbf{OS-Atlas-Web}, the synthesized data from OS-Atlas dataset, derived from FineWeb~\cite{penedo2024fineweb}; (2) \textbf{\ourmethod{}}, 500k instructions sampled from the web portion of our final training dataset; (3) \textbf{-~Inst. Synth.}, a variant that skips the instruction synthesis step and uses referring expressions as instructions; and (4) \textbf{-~GPT-4o}, a configuration using only raw element attributes as instructions.  
We sample 500k instruction from above configurations and fine-tuned the same InternVL2-4B model separately on each dataset and evaluated the trained models on the ScreenSpot benchmark. As shown in Table~\ref{tab:curation_ablation}, the results demonstrate that our dataset yields significant performance improvement compared to OS-Atlas-Web. While this advantage diminishes with OS-Atlas larger training data in Table~\ref{tab:screenspot}, it also demonstrate the effectiveness of our curated data. Furthermore, the performance degradation observed when eliminating instruction synthesis and GPT-4o components provides evidence for our \ourmethod{} design.

\begin{table}[t]
    \centering
    \centering
    \resizebox{\columnwidth}{!}
    {
    \begin{tabular}{lcccccc}
    \toprule
    \textbf{Setting} & \textbf{OS} & \textbf{Office} & \textbf{Daily} & \textbf{Profess.} & \textbf{Workflow} & \textbf{Overall} \\ 
    \midrule
    Screenshot only$^*$    & 8.3   & 2.6   & 6.5   & 0   & 2.1  & 3.6 \\ 
    +OS-Atlas & 30.4 & 2.7 & 11.7 & 16.3 & 2.2 & 8.5\\
    +\ourmodel{} & 26.1 & 5.2 & 23.4 & 18.8 & 2.8 & 12.0\\
    \bottomrule
    \end{tabular}
    }
    \caption{Task success rate on OSWorld. $^*$Report as the performance reproduced by us. In this experiment, the environment and settings are consistent.}
    \vspace{-16pt}
    \label{tab:osworld}
    \end{table}

\paragraph{Failure case analysis.}

Figure~\ref{failurecase} illustrates common errors of \ourmodel{} on \ourbench{}, including: (1) failure to recognize icons without text due to limited knowledge, (2) incorrect positioning of elements within rows or columns, (3) misinterpretation of spatial relationships, (4) misunderstanding hierarchical relationships, and (5) misclassifying element types, such as confusing checkboxes with adjacent text.

\subsection{Agent Task Online Evaluation}

\paragraph{OSWorld.} We evaluate our model on OSWorld~\cite{osworld}, a live benchmark built in real computer environment for GUI agents. OSWorld consists of 369 open-ended computer tasks across various operating systems, including Ubuntu, Windows, and macOS. 
Following its original screenshot-only design, we use GPT-4o as the planner, providing only screenshots as the observation space. For every step, the planner is given the user task and the screenshot of current stage to predict the next action to take in next step.
Then for the screenshot and the predicted action, we leverage VLMs to provide corresponding element coordinates to compare the effect of different VLMs. 
For screenshot only baseline, we use GPT-4o to directly generate element coordinates.
The prompt used in evaluation is provided in Appendix~\ref{synthesis_prompt}.
The result comparison is shown in Table~\ref{tab:osworld}. 
The results show that \ourmodel{} helps to address GPT-4o's limitations in grounding and shows competitive performance, which prove our data synthesis pipeline can benefit the GUI agent development.

\section{Conclusion}
This work tries to identify and address critical challenges in GUI instruction grounding, such as element-to-screen ratio, unbalanced element types, and implicit instructions, through the introduction of a large-scale data synthesis pipeline, \textit{\ourmethod}. Moreover, we propose a new GUI instruction grounding benchmark, \textit{\ourbench}, which overcomes the limitations of existing benchmarks by incorporating diverse annotation aspects. Our model, trained on the synthesized data, outperforms the previous state-of-the-art grounding models with less data and a smaller model size. We hope this work provides valuable insights and tools for the future development of GUI instruction grounding.

\section*{Limitations}
While the introduction of a large-scale data synthesis pipeline (\textit{\ourmethod}) has significantly improved the generation of extensive instruction datasets, there is still potential for further scaling the data size and model size. Increasing the dataset size could enhance the robustness and generalizability of the model, addressing more diverse and complex GUI scenarios. Another limitation is this work only focuses on English instruction, we hope to further apply our data synthesis pipeline on other lanugages.

\section*{Ethical considerations}
Automated GUI agents can liberate human from repetitive tasks in digital devices but also could be used to launch robot attacks, such as brute force login attempts, automated spam, or denial-of-service attacks, by interacting with web interfaces at a speed and scale beyond human capability.
They might be used to automate phishing attacks, where the agent mimics legitimate user interactions to steal sensitive information.
It necessitates the maintaining transparency about the capabilities and limitations of GUI agents.

\bibliography{custom}

\appendix

\section*{Appendix}

The writing of this paper was limitedly assisted by AI, with refinement focused on language and style, and no original ideas generated by the AI.

\section{Data Statistics}
\label{appendix:statistic}
We present a detailed summary of the statistics for the complete training dataset we used in Table~\ref{tab:data_statistics}. All the external dataset used are publicly available with Apache 2.0 License.
\begin{table}[h]
\centering
\resizebox{\columnwidth}{!}
{
\begin{tabular}{lccccccc}
\toprule
\textbf{Dataset} & \textbf{Platform} & \textbf{\#Screenshots}  & \textbf{\#Instructions}  \\
\midrule
\textit{\ourmethod}-Web &  Web & 1,536,200 & 9,097,736 \\
\textit{\ourmethod}-Desktop  & Desktop & 14,087 & 334,397 \\
\textit{\ourmethod}-AndroidControl   & Mobile & 40,199 & 109,126 \\
% WindowsIcon & Desktop & 12,050 & 110,990 \\
MOTIF\cite{burns2022motifvln} & Mobile & 30,699 & 320,219 \\
WidgetCaption\cite{li2020widget} & Mobile & 14,409 & 38,103 \\
\midrule
Total & - & 1,635,594 & 9,899,581 \\
\bottomrule
\end{tabular}
}
\caption{Grounding training datasets statistics.}
\label{tab:data_statistics}
\end{table}

We further analyzed the statistics of the synthetic data. In our synthetic data, the proportion of non-text elements is 23.0\%. For comparison, we randomly sampled 100 elements from the SeeClick web data, manually labeled the non-text elements, and calculated their proportion. In SeeClick, non-text elements only account for 8.7\%. Our dataset exhibits a more balanced distribution across different element types. 

Additionally, we randomly sampled 1,000 elements from landscape screenshots in both datasets and calculated their element-to-screen ratio. The distribution in shown in Table~\ref{tab:element_ratio_comparison}. Compared to SeeClick, we constructed our dataset with more data for smaller and more challenging targets, in order to optimize the model's performance in high-resolution real-world scenarios.

\begin{table}[h]
\centering
\resizebox{\columnwidth}{!}
{
\begin{tabular}{c|cc}
\toprule
\textbf{\makecell{Element-to-screen\\Ratio Range}} & \textbf{\makecell{Proportion (\%)\\(SeeClick)}} & \textbf{\makecell{Proportion (\%)\\(Ours)}} \\
\midrule
0.00 - 0.02 & 11.49 & 36.92 \\
0.02 - 0.04 & 43.30 & 40.43 \\
0.04 - 1.00 & 45.21 & 22.65 \\
\bottomrule
\end{tabular}
}
\caption{Element-to-screen ratio comparison between SeeClick web dataset and our synthetic web dataset.}
\label{tab:element_ratio_comparison}
\end{table}

\section{Training Details}
\label{appendix:training_details}
\paragraph{\ourmodel{}-7B.}
To maintain consistency with the Qwen2-VL training data, we converted the format of the bounding boxes to \texttt{<|box\_start|>(x1,y1), (x2,y2)<|box\_end|>}, where (x1, y1) and (x2, y2) represent the coordinates of the upper left and bottom right corners of the box, normalized within the range [0, 1000). \texttt{<|box\_start|>} and \texttt{<|box\_end|>} are treated as special tokens. We pack the samples corresponding to the same image into a single conversation, with no more than 15 samples per conversation. To achieve better performance on high-resolution images, we set the \texttt{max\_pixels} to 1500*1500.
We perform full-parameter fine-tuning on our dataset. The whole training process costs around 60 hours on 16 A100 GPU cards.

\paragraph{\ourmodel{}-4B.}
Following the pre-training setting, we convert bounding boxes into the format \texttt{<box>}[[x1, y1, x2, y2]]\texttt{</box>}, where [x1, y1, x2, y2] are the original coordinates on the image without normalization. \texttt{<box>} and \texttt{</box>} are treated as special tokens. Through experiments we found that packing the data caused a performance degradation of InternVL2-4B, so we did not perform packing. 

We apply the 2-D bounding box tile tag proposed in ~\cite{dai2024nvlm} to improve the model's visual ability.
Since InternVL-2 leverages Dynamic Aspect Ratio to handle images of varying resolutions, during both training and inference, we transform the original coordinates to align with the resized image size within the model. 
To accommodate the high-resolution nature of the UI interface, we set \texttt{patch\_num} to 12, which means that each image is divided into 12 tiles of 448x448 pixels. 

We perform full-parameter fine-tuning on the language model on our dataset. The whole training process costs around 90 hours on 64 A100 GPU cards.

\section{Details of \textit{\ourbench} Annotation}

\begin{table}[h]
\centering
\resizebox{0.8\columnwidth}{!}
{

\begin{tabular}{llc}
\toprule
\textbf{Error Category} & \textbf{Severity} & \textbf{Proportion (\%)}  \\
\midrule
Element Box & Serious & 9.7 \\
            & Slight & 24.1 \\
\midrule
Element Instruction & Serious  & 17.3 \\
                   & Slight & 7.9 \\
\bottomrule
\end{tabular}
}
\caption{Distribution of error types of \textit{\ourmethod} generated data in \textit{\ourbench}}
\label{tab:bench_quality}
\end{table}

\label{appendix:benchmark}

To reduce the workload of manual annotation, we adopt our \textit{\ourmethod} pipeline on multiple platforms to construct the \textit{\ourbench}. We do balanced sampling for every element type, resulting in 1,987 elements.
Then we manually annotated each data, including box quality, instruction quality and instruction type. 
For box quality, we consider bounding boxes that do not enclose any interactive UI elements as serious errors. 
Boxes that contain the correct element but are slightly misaligned or have inaccurate boundaries are labeled as slight errors. 
We retained data with slight errors in our benchmark and manually adjusted the boxes.
For instruction quality, we consider instructions that are completely unrelated to any elements to be serious errors. Instructions that are unclear or ambiguous are labeled as slight errors. Similarly, we modified the slightly erroneous instructions and retained them.
In addition, we annotated the instruction type for each data. Instructions that directly refer to an element are labeled as explicit, while those describe the element indirectly by position, appearance, or other references are implicit. 

\begin{figure*}
    \centering
    \includegraphics[width=0.85\textwidth]{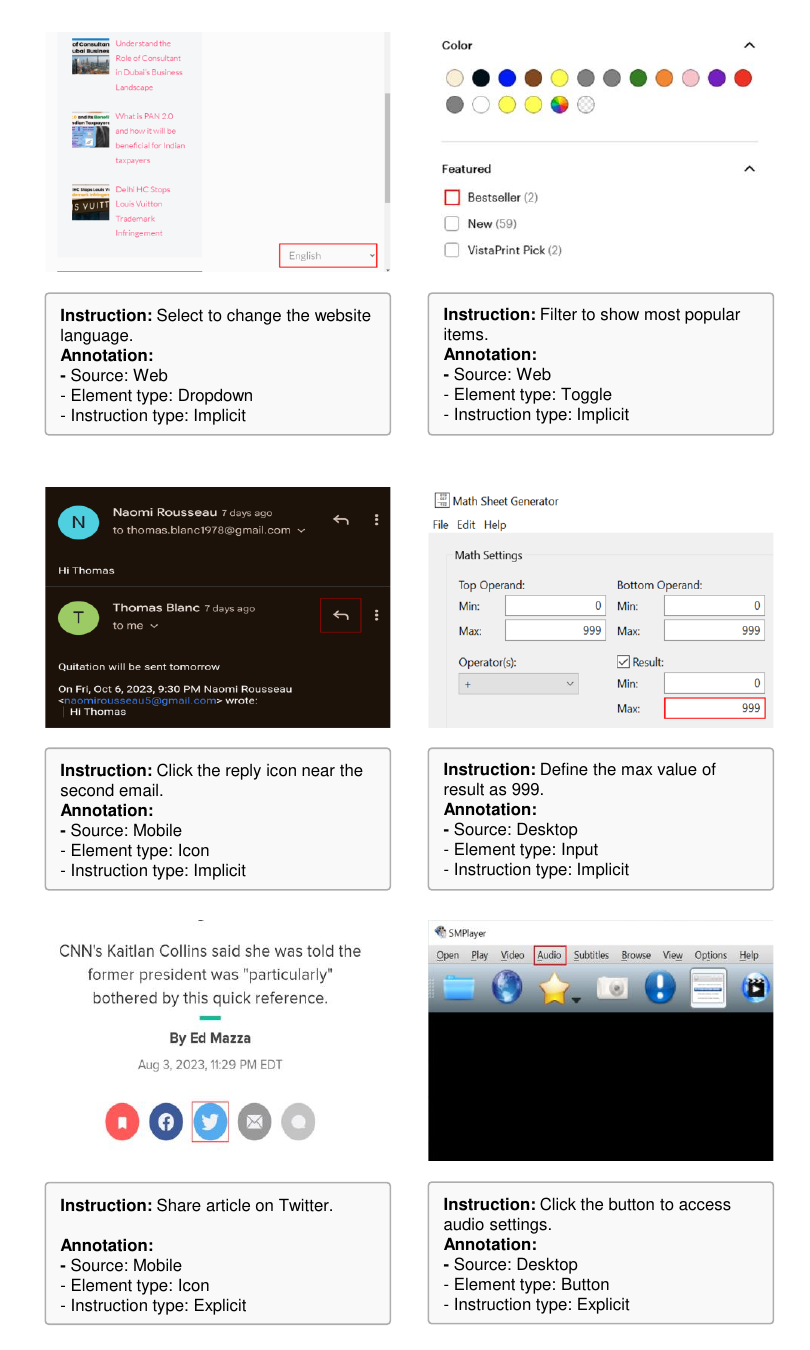}
    \caption{Examples in our benchmark \ourbench{}. Red boxes denote the target elements. In the examples shown in this figure, we cropped only a portion of the original image.}
    \vspace{-8pt}
    \label{fig:benchmark_case}
\end{figure*}

After manual annotation, we removed data with serious errors, resulting in a benchmark containing 1,477 element-instruction pairs. The distribution of error types of data are shown in Table~\ref{tab:bench_quality}. We present some examples of our benchmark in Figure~\ref{fig:benchmark_case}. All annotators are authors of this work and they all are Asian using English.

\begin{figure}
    \centering
    \includegraphics[width=0.89\linewidth]{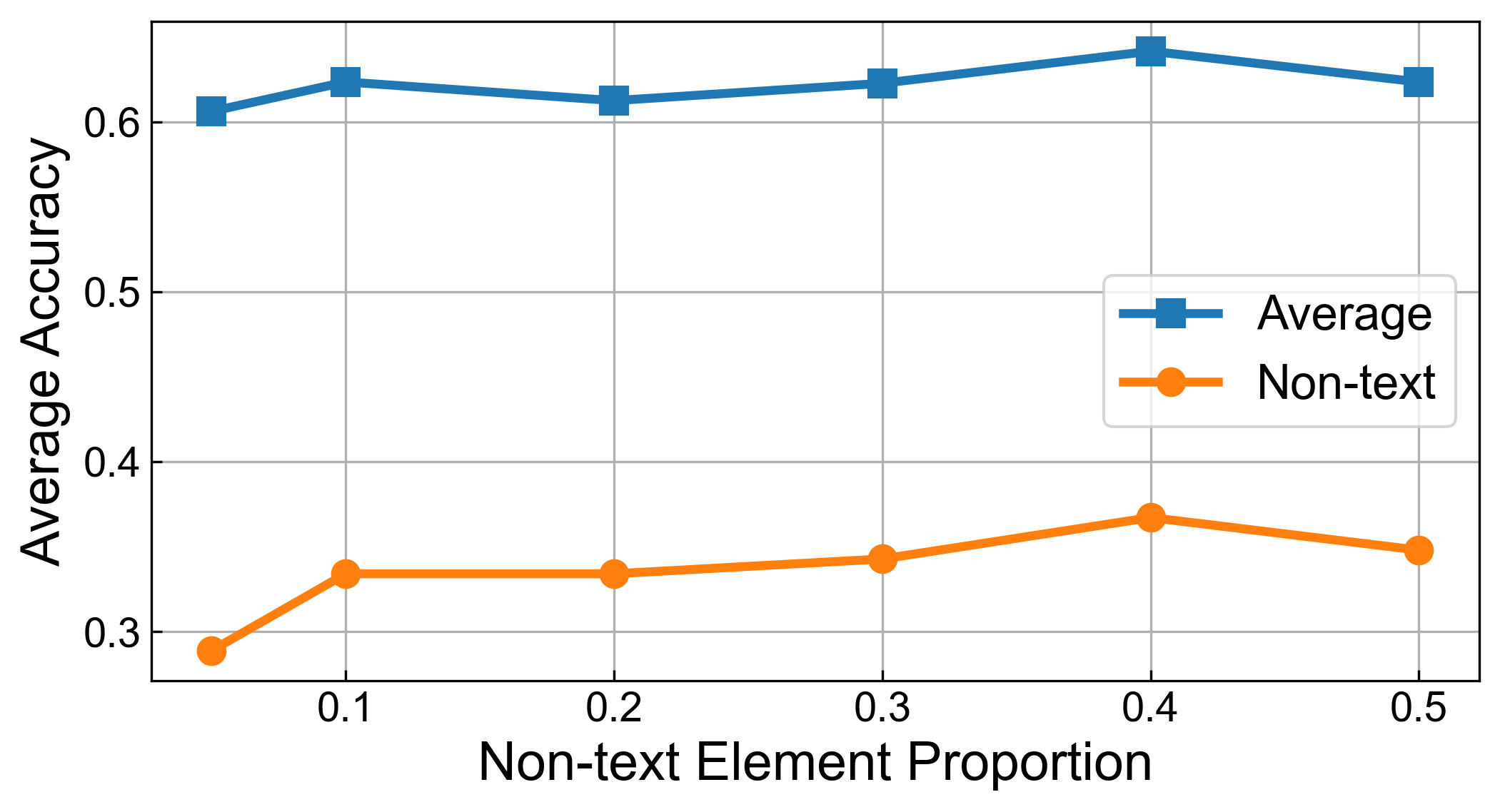}
    \caption{Performance of model trained on 500k UI elements with different proportions of non-text elements in training dataset. The Average represents accuracy on the entire ScreenSpot, while the Non-text represents accuracy specifically on icons/widgets in ScreenSpot.}
    \vspace{-8pt}
    \label{fig:nontext_ratio_ablation}
\end{figure}

\begin{table*}[h]
\centering
\resizebox{\textwidth}{!}
{
\centering
\begin{tabular}{lllcccccccc}\toprule
\multirow{2}{*}{\textbf{Model}} &\multirow{2}{*}{\textbf{Size}} &\multirow{2}{*}{\textbf{\#Train}} &\multicolumn{2}{c}{\textbf{Mobile}} &\multicolumn{2}{c}{\textbf{Desktop}} &\multicolumn{2}{c}{\textbf{Web}} &\multirow{2}{*}{\textbf{Avg.}} \\
\cmidrule(lr){4-5} \cmidrule(lr){6-7} \cmidrule(lr){8-9}
& & &Text &Icon/Widget &Text &Icon/Widget &Text &Icon/Widget & \\\midrule
InternVL2-4B & 4B & - & 9.2 & 4.8 & 4.6 & 4.3 & 0.9& 0.1 & 4.2\\
Qwen2-VL-7B & 7B & - & 61.3 & 39.3 & 52.0 & 45.0 & 33.0 & 21.8 & 42.6\\
OmniParser & - & - & 87.4 & 67.9 & 93.2 & 60.9 & 77.8 & 48.3 & 73.9\\
CogAgent &17B & 37M$^{*}$ & 67.0&24.0& 74.2 &20.0 &70.4 &28.6 &47.4 \\
SeeClick & 9.6B & 1.0M & {78.0} &52.0 &72.2 &30.0 &55.7 &32.5 & 55.8 \\
UGround & 7B & 9.7M & 82.8 & 60.3 & 82.5 & 63.6 & 80.4 & 70.4 & 74.1\\
ShowUI & 2B & 2.7M & 92.3 & 75.5 & 76.3 & 61.1 & 81.7 & 63.6 & 76.8 \\
OS-Atlas-4B & 4B & 13.6M & 85.7 & 58.5 & 72.2 & 45.7 & 82.6 & 63.1 & 70.1\\
OS-Atlas-7B & 7B & 13.6M & 93.0 & 72.9 & 91.8 & 62.9 & 90.9 & 74.3 & 82.5 \\
\midrule
\ourmodel{}-4B & 4B & 9.9M & 87.6 & 49.8 & 87.6 & 45.7 & 85.7 & 54.4 & 70.4 \\
\ourmodel{}-7B & 7B & 9.9M & 94.1 & 77.3 & 90.7 & 71.4 & 87.4 & 67.5 & 82.5\\
\bottomrule
\end{tabular}
}
\caption{ Full results on ScreenSpot. $^{*}$Train data for CogAgent includes both pre-training and multi-task finetuning data.}
\label{tab:screenspot_full}
\end{table*}

\begin{table*}[h]
\centering
\resizebox{\textwidth}{!}
{
\begin{tabular}{lccccccc}
\toprule
\textbf{Model} & \textbf{Development} & \textbf{Creative} & \textbf{CAD} & \textbf{Scientific} & \textbf{Office} & \textbf{Operating Systems} & \textbf{Overall Avg} \\
\midrule
InternVL2-4B & 0.3 & 0.0 & 0.0 & 0.4 & 0.4 & 0.5 & 0.3\\
Qwen2-VL-7B & 1.3 & 0.9 & 0.4 & 3.5 & 3.0 & 0.5 & 1.6 \\
OmniParser+GPT-4o & 13.7 & 1.5 & 7.7 & 9.4 & 14.3 & 4.6 & 8.3\\

SeeClick & 0.3 & 0.6 & 1.9 & 2.0 & 0.9 & 1.5 & 1.1\\
UGround & 14.7 & 17.0 & 11.1 & 19.3 & 27.0 & 9.7 & 16.5\\
ShowUI & 9.4 & 5.3 & 1.9 & 10.6 & 13.5 & 6.6 & 7.7 \\
OS-Atlas-4B & 3.7 & 2.3 & 1.5 & 7.5 & 4.8 & 3.1 & 3.7\\
OS-Atlas-7B & 17.7 & 17.9 & 10.3 & 24.4 & 27.4 & 16.8 & 18.9 \\
\midrule
\ourmodel{}-4B & 12.7 & 12.6 & 5.7 & 13.4 & 15.2 & 14.3 & 12.2\\
\ourmodel{}-7B & 24.1 & 23.5 & 9.2 & 27.6 & 38.3 & 19.9 & 23.6\\
\bottomrule
\end{tabular}
}
\caption{ Full results on ScreenSpot-Pro(simple view).}
\label{tab:screenspotpro_full}
\end{table*}

\section{Ablation on Non-Text Element Proportion}
In unfiltered data, the quantity of non-text elements is usually smaller than that of text elements. However, identifying non-text elements is much more challenging. When computational resources are limited, adjusting the proportion of non-text elements in the total dataset can help to improve model performance. 
We investigate the impact of the proportion of non-text elements on accuracy. We use InternVL2-4B as the base model, and sample 500K elements from 250K images for each proportion. The results are shown in Figure~\ref{fig:nontext_ratio_ablation}. Within a certain range, increasing the proportion of non-text elements can effectively improve the accuracy of non-text elements.

\section{Related literature discussion}

GUI agent is a multimodal agent designed for GUI scenarios, which requires multi-step planning and reasoning ability as general language agents~\cite{sumerscognitive, yang2024swe, xi2025rise} do. However, due to the incompleteness and opacity of UI metadata, the information within a GUI cannot be fully represented through pure language. It is necessary to leverage the multimodal perception capabilities of models to understand the current task state via GUI screenshots. During action execution, multimodal capabilities are also required to extract the coordinates of interactive elements from the current GUI screenshot, with GUI grounding playing a crucial role. Nevertheless, current general-purpose VLM, especially open-sourced leading MLLMs such as InternVL2~\cite{chen2023internvl} and Qwen2-VL~\cite{qwen2vl}, exhibit relatively limited perception and planning capabilities for GUI tasks. 
Hence, the collection of corresponding training data is essential and warrants community attention.

\section{Prompt templates}
\label{synthesis_prompt}
We display the prompts for the instruction synthesis in \ourmethod{}, prompts for OmniParser evaluation and prompts given to GPT-4o in OSWorld evaluation as follows.

\section{Insights on failure cases in \ourbench}
\label{appendix:failure_analysis}
Here we provide preliminary analysis about the typical failure cases in \ourbench{} separately:

\paragraph{Counting error and spatial relationship misunderstanding.} The spatial understanding has always been two of the challenges faced by MLLMs. Due to the bias in MLLM pre-training data, models are typically better at detecting whether an object exists in the image but struggle with identifying its quantity and location. We believe this issue can be mitigated by adding data with instructions related to quantities and spatial relationships during training. In fact, through our synthetic data pipeline, our model has already made considerable progress compared to the baseline models in addressing these issues.

\paragraph{Element type misclassification.} In UI environment, the interactive region of different element types naturally varies. In general, large-scale training data contains elements with the same contents but different types, with some types of elements dominating in quantity(e.g., text). As a result, the model is more likely to assume that elements belong to these dominate types, leading to incorrect judgements on their interactive regions. 

\paragraph{Lack of external knowledge.} While some common cases can be addressed by scaling up the range of training data, it is important to note that covering all icons from a variety of specialized software in the training data is challenging. We believe that for specific software, creating explanatory documentation for its icons and providing it as contextual reference for the model could be a potential solution in the future.

\paragraph{UI hierarchical misunderstanding.} Most current grounding models directly output bounding boxes, lacking an analysis and reasoning progress regarding UI hierarchies. In fact, when dealing with grounding tasks that need page semantic understanding, the model sometimes needs to perform multi-step reasoning. In the example of Figure~\ref{failurecase} Task 4, the model needs to first locate the login module and then identify the e-mail input field. A potential mitigation strategy is to use CoT reasoning.

\section{GPT-4o response example in Instruction Synthesis}
\label{appedix:is}
For the parameter-based instruction synthesis, we do not simply repeat the already generated action parameters and assemble them to instruction. Instead, we use them as reference to guide the final generation of first-person perspective instructions. We here provide examples of the action parameter and corresponding synthesized instructions sampled from generated dataset. Rather than merely rigidly assembling the given parameters, the synthesized instructions generated from GPT-4o demonstrate naturalness and fluency, exhibiting newly generated content that diverges from the provided action parameters.

\onecolumn
\begin{tcolorbox}[colback=white, % background color
    colframe=gray!50!black, % frame color
    coltitle=black, % title color
    title=\textbf{Prompts for instruction synthesis in \textit{\ourmethod}}, % box title
    title after break=\textbf{(cont.) Prompts for instruction synthesis in \textit{\ourmethod}}, 
    fonttitle=\bfseries\large, % title font
    colbacktitle=white!80!gray, coltitle=black,
    left = 1mm,right = 1mm,top = 1mm,bottom = 1mm,
    enhanced, % enable advanced features
    skin first=enhanced,
    skin middle=enhanced,
    skin last=enhanced,
    breakable=true,
    after skip=0pt,
    ]
\footnotesize
{
\textbf{Step1}:
\\\\
You are a screen UI expert. Here is a UI screenshot image with highlighted bounding boxes and corresponding labeled ID overlayed on bottom of them, and here is a list of corresponding UI element (icon/button/inputfield) box description within these bounding boxes. You should first ensure you understand the screenshot's status and every annotated UI element bounding box on it. Then output the updated element list with below template as JSON format.\\
\\
NOTE: Ensure all referring exprression should UNIQUELY correspond to the element when generating them.

Here is the output template:
\begin{Verbatim}[breaklines=true]
{
    "elements": [
        {
            "id": "<copy corresponding labeled ID on the bottom of element>", 
            "shortDescription": "<copy from given input list>",
            "fullDescription": "<a comprehensive description that explains the function of the element and the expected outcome when interacting with it>",
            "explicitRefer": "<a referring expression that explicitly refers to the element, from a computer user's first perspective>", 
            "implicitReferByElementFunction": "<a referring expression that does NOT explicitly refer to this element content or obvious visual feature, but implicitly refers to it by its function in the whole page or expected outcome after interacting with it>",
            "implicitReferByNearElement": "<a referring expression that does NOT explicitly refer to this element content or obvious visual feature, but implicitly refers to it by its relationship with near elements or emphasizes it from similar elements by spatial order>"
        }
    ]
}
\end{Verbatim}

\textbf{Step2}: 

You are a computer expert user. You will be given a UI element list from a UI screenshot which includes elements and their referring expressions as shown in the following input template. You should simulate a user using the UI screenshot, generate possible action type and action content, then generate instructions for every element according to the given element referring expressions, as shown in the following output template. Return as JSON format.\\
\\
NOTE:
\\
1. For inputfield element type, the generated instruction should contain the possible input content from a computer user's first perspective. For example, "fill James as last name", "enter Boeing747 in search field", "select article in recent six months".

2. Ensure all instruction should UNIQUELY correspond to the element when generating them.

Here is the input template:
\begin{Verbatim}[breaklines=true]
{
    "elements": [
        {
            "id": "<copy corresponding labeled ID on the bottom of element>", 
            "shortDescription": "<a short description about the element, including element type and element content>",
            "fullDescription": "<a comprehensive description that explains the function of the element and the expected outcome when interacting with it>",
            "explicitRefer": "<a referring expression that explicitly refers to the element, from a computer user's first perspective>", 
            "implicitReferByElementFunction": "<a referring expression that does NOT explicitly refer to this element content or obvious visual feature, but implicitly refers to it by its function in the whole page or expected outcome after interacting with it>",
            "implicitReferByNearElement": "<a referring expression that does NOT explicitly refer to this element content or obvious visual feature, but implicitly refers to it by its relationship with near elements or emphasizes it from similar elements by spatial order>"
        }
    ]
}
\end{Verbatim}

Here is the output template:

\begin{Verbatim}[breaklines=true]
{
    "elements": [
        {
            "id": "<copy corresponding labeled ID on the bottom of element>", 
            "shortDescription": "<copy from given input list>",
            "instructionArgs": {
                "actionType": "<choose appropriate action type>",
                "actionContentDescription": "<describe what is the possible action content about. if CLICK: leave empty string. if TYPE: specific input content which is appropriate for current UI screenshot. if Toggle, Checkbox and Switch, ON or OFF>",
                "actionContent": "<According to `actionContentDescription` to fill the specific content, if CLICK: leave empty string>"
            },
            "convertedUserInstructionByElementFunction": "<including above `actionContent` if not empty, convert above `implicitReferByElementFunction` into a computer user's first perspective instruction, concise and less than 15 words>",
            "convertedUserInstructionByNearElement": "<including above `actionContent` if not empty, convert above `implicitReferByNearElement` into a computer user's first perspective instruction, concise and less than 15 words>"
        }
    ]
}
\end{Verbatim}
}
\end{tcolorbox}

\begin{tcolorbox}[colback=white, % background color
    colframe=gray!50!black, % frame color
    coltitle=black, % title color
    title=\textbf{Prompts for OmniParser Evaluation}, % box title
    fonttitle=\bfseries\large, % title font
    colbacktitle=white!80!gray, coltitle=black,
    left = 1mm,right = 1mm,top = 1mm,bottom = 1mm,
    enhanced, % enable advanced features
    breakable=true,
    after skip=0pt,
    ]
\footnotesize
{
Here is a UI screenshot image with bounding boxes and corresponding labeled ID overlayed on top of it, and here is a list of icon/text box description: \texttt{\{parsed\_local\_semantics\}}. 
\\
Your task is \texttt{\{task\}}. Which bounding box label you should operate on? Give a brief analysis, then put your answer in the format of: \\Box with label ID: [xx] 
}
\end{tcolorbox}

\begin{tcolorbox}[colback=white, % background color
    colframe=gray!50!black, % frame color
    coltitle=black, % title color
    title=\textbf{Prompts for GPT-4o in OSWorld}, % box title
    title after break=\textbf{(cont.) Prompts for GPT-4o in OSWorld},
    fonttitle=\bfseries\large, % title font
    colbacktitle=white!80!gray, coltitle=black,
    left = 1mm,right = 1mm,top = 1mm,bottom = 1mm,
    enhanced, % enable advanced features
    skin first=enhanced,
    skin middle=enhanced,
    skin last=enhanced,
    breakable=true,
    ]
\footnotesize
{
You are an agent which follow my instruction and perform desktop computer tasks as instructed. \\
You have good knowledge of computer and good internet connection and assume your code will run on a computer for controlling the mouse and keyboard.
For each step, you will get an observation of an image, which is the screenshot of the computer screen and you will predict the action of the computer based on the image. \\
\\
You are required to use `pyautogui` to perform the action grounded to the observation, but DONOT use the `pyautogui.locateCenterOnScreen` function to locate the element you want to operate with since we have no image of the element you want to operate with. DONOT USE `pyautogui.screenshot()` to make screenshot. \\
Return exactly ONE line of python code to perform the action each time. At each step, you MUST generate the corresponding instruction to the code before a \# in a comment (example: \# Click \textbackslash "Yes, I trust the authors\textbackslash " button\textbackslash n  pyautogui.click(x=0, y=0, duration=1))  \\
\\
You need to to specify the coordinates of by yourself based on your observation of current observation, but you should be careful to ensure that the coordinates are correct.
You ONLY need to return the code inside a code block, like this: \\
```python \\
\# your code here \\
``` \\
Specially, it is also allowed to return the following special code:
When you think you have to wait for some time, return ```WAIT```;
When you think the task can not be done, return ```FAIL```, don't easily say ```FAIL```, try your best to do the task;
When you think the task is done, return ```DONE```.\\
\\
Here are some guidelines for you:\\
1. Remember to generate the corresponding instruction to the code before a \# in a comment.\\
2. If a click action is needed, use only the following functions: pyautogui.click, pyautogui.rightClick or pyautogui.doubleClick.\\
3. Return ```Done``` when you think the task is done. Return ```Fail``` when you think the task can not be done.\\
4. If there are many times in the history of the same repeated action, first try to change the expression of the instruction, and if it still fails, try something else.\\
\\
My computer's password is 'password', feel free to use it when you need sudo rights.
First give the current screenshot and previous things we did a short reflection, then RETURN ME THE CODE OR SPECIAL CODE I ASKED FOR. NEVER EVER RETURN ME ANYTHING ELSE.\\
\\
Please generate the next move according to the UI screenshot, instruction and previous actions.\\
\\
Instruction: \texttt{\{instruction\}}.

Previous actions:
\texttt{\{previous\_actions\}}
}
\end{tcolorbox}

\begin{tcolorbox}[colback=white, % background color
    colframe=gray!50!black, % frame color
    coltitle=black, % title color
    title=\textbf{GPT-4o response example in \ourmethod}, % box title
    title after break=\textbf{(cont.) GPT-4o response example in \ourmethod},
    fonttitle=\bfseries\large, % title font
    colbacktitle=white!80!gray, coltitle=black,
    left = 1mm,right = 1mm,top = 1mm,bottom = 1mm,
    enhanced, % enable advanced features
    skin first=enhanced,
    skin middle=enhanced,
    skin last=enhanced,
    breakable=true,
    ]
\footnotesize
{
\begin{Verbatim}[breaklines=true]
[
    {
        "instructionArgs": {
            "actionType": "TYPE",
            "actionContent": "Gangsta-Groove"
        },
        "convertedUserInstructionByElementFunction": "Enter 'Gangsta-Groove' in the search bar",
        "convertedUserInstructionByNearElement": "Type 'Gangsta-Groove' in the search field above the article",
    },
    {
        "instructionArgs": {
            "actionType": "CLICK",
            "actionContent": ""
        },
        "convertedUserInstructionByElementFunction": "Review price based on net area.",
        "convertedUserInstructionByNearElement": "Check near rent/total price toggle.",
    },
    {
        "instructionArgs": {
            "actionType": "SELECT",
            "actionContent": "preferred language"
        },
        "convertedUserInstructionByElementFunction": "Select your preferred language",
        "convertedUserInstructionByNearElement": "Choose a language next to Contact Us",
    },
    {
        "instructionArgs": {
            "actionType": "CLICK",
            "actionContent": ""
        },
        "convertedUserInstructionByElementFunction": "Click to learn more about us",
        "convertedUserInstructionByNearElement": "Click the second button on the menu"
    },
    {
        "instructionArgs": {
            "actionType": "CLICK",
            "actionContent": ""
        },
        "convertedUserInstructionByElementFunction": "Return to the top of the page",
        "convertedUserInstructionByNearElement": "Click the bottom right corner icon",
    }
]
\end{Verbatim}
}
\end{tcolorbox}

\twocolumn

\end{document}